\documentclass{jfm}
\usepackage{graphicx} 
\usepackage{natbib} 
\usepackage{epstopdf}
\usepackage[font=small]{caption}
\usepackage[labelformat = empty,position=top]{subcaption}
\usepackage[export]{adjustbox} 
\usepackage{textcomp}
\usepackage{metre}
\ifCUPmtlplainloaded \else
  \checkfont{eurm10}
  \iffontfound
    \IfFileExists{upmath.sty}
      {\typeout{^^JFound AMS Euler Roman fonts on the system,
                   using the 'upmath' package.^^J}%
       \usepackage{upmath}}
      {\typeout{^^JFound AMS Euler Roman fonts on the system, but you
                   dont seem to have the}%
       \typeout{'upmath' package installed. JFM.cls can take advantage
                 of these fonts,^^Jif you use 'upmath' package.^^J}%
      }
  \else
  \fi
\fi
\ifCUPmtlplainloaded \else
  \checkfont{msam10}
  \iffontfound
    \IfFileExists{amssymb.sty}
      {\typeout{^^JFound AMS Symbol fonts on the system, using the
                'amssymb' package.^^J}%
       \usepackage{amssymb}%
       \let\le=\leqslant  
       \let\ge=\geqslant  
      }{}
  \fi
\fi

\ifCUPmtlplainloaded \else
  \IfFileExists{amsbsy.sty}
    {\typeout{^^JFound the 'amsbsy' package on the system, using it.^^J}%
     \usepackage{amsbsy}}
    {}
\fi

\newcommand\Pran{\mbox{\textit{Pr}}} 
\newsavebox{\astrutbox} \sbox{\astrutbox}{\rule[-5pt]{0pt}{20pt}}

\newcommand\p{\ensuremath{\partial}}

 \title[A criterion to detect line
plumes from velocity fields]{A criterion to detect line plumes from
  velocity fields in turbulent convection} \author[Vipin K. and
Puthenveettil B. A.]%
{Koothur Vipin\thanks{Email address for correspondence:
    vipink159@gmail.com} and Baburaj A. Puthenveettil}
\affiliation{Department of Applied Mechanics, Indian Institute of
  Technology Madras, Chennai 600036, India}
\pubyear{2015}
\volume{x}
\pagerange{x--y}
\date{?; revised ?; accepted ?. - To be entered by editorial office}
\begin{document}
\maketitle
\maketitle
\begin{abstract} 
  We present a simple, new criterion to extract line plumes from the
  velocity fields, without using the temperature field, in a
  horizontal plane close to the plate in turbulent convection. The
  existing coherent structure detection criteria from velocity fields,
  proposed for shear driven wall turbulence, are first shown to be
  inadequate for turbulent convection. Based on physical arguments, we
  then propose that the negative values of
  $\overline{\nabla_H}\cdot\overline{u}$, the horizontal divergence of
  the horizontal velocity field extracts the plume regions from the
  velocity field. The $\overline{\nabla_H}\cdot\overline{u}$ criterion
  is then shown to predict the total length of plumes correctly over
  a 
  Rayleigh number range of $7.7\times 10^5\le Ra_w\le 1.43\times 10^9$
  and a Prandtl number range of $3.96\le Pr\le
  5.3$. 
  The thickness of the plume region predicted by the present criterion
  is approximately same as that obtained by a threshold of RMS
  temperature fluctuations, $\sqrt{T'^2}=0.2 \Delta T$, where
  $\Delta T$ is the temperature drop across the fluid layer.
\end{abstract}
\begin{keywords}
\end{keywords}
\section{Introduction}
\label{sec:introduction}
Persistent line plumes are the dominant coherent structures on hot
surfaces in turbulent convection~\cite*[][]{Theerthan94}. The spatial
distribution of such line plumes on the hot plate have been studied
mostly by observing concentration of dyes
\cite*[][]{Sparrow69,adr,Theerthan98,mine,Puthenveettil05a,
  Puthenveettil11,Guna14} or by visualizing temperature sensitive
tracers like liquid crystals~\cite*[][]{Zhou02}. The plume structure
has also been obtained by~\cite{Shishkina08} using criteria like
conditional averaged values on temperature of vertical velocity,
absolute value of horizontal velocity, heat flux, thermal dissipation
rates and the vertical and horizontal vorticity components. As per
them, a threshold of conditional averaged thermal dissipation rates on
temperature are the best for extracting plumes from numerical
simulations. At a location, plumes have been detected temporally by
analysing the temperature signals~\cite*[][]{Belmonte96} as well as
the conditional averaged velocity on the
temperature~\cite*[][]{Ching04}. To apply all the above methods and
obtain the spatial pattern of line plumes, one needs scalar fields
like temperature or concentration, often along with the 3D velocity
fields. In experimental studies, only the velocity fields are
available, as in most particle imaging velocimetry (PIV) studies;
these methods are hence unsuitable to detect the spatial pattern of
line plumes.

Detection of coherent structures using velocity fields alone have been
extensively studied in wall flows where turbulence is solely produced
by shear (\cite{Tardu}). Some of the criteria used in these flows
include $Q$-criterion~\cite*[][]{hunt88:_eddies}, $\lambda_2$
criterion~\cite*[][]{Jeong95} and swirling strength ($\lambda_{ci}$)
method~\cite*[][]{Zhou99}. All these methods are based on the local
analysis of the velocity gradient tensor
$\overline{\nabla} \overline{u}$, often using all its nine components.
To get all the components of $\overline{\nabla} \overline{u}$, one
needs the three velocity components and the three gradients of each of
the velocity components, all of these are most often not available in
experimental studies. For most stereo PIV studies, the three velocity
components and their horizontal gradients in a plane are only
available; the data is further limited to the two horizontal
components of velocities and their horizontal gradients for 2D PIV. In
such cases, to detect coherent structures in wall shear turbulence,
the above criteria are calculated with the unavailable components of
$\overline{\nabla} \overline{u}$, often the components that use the
vertical gradients of velocity, set to zero; the approximation seems
to work fairly well for wall shear flows. Such an approximation
however appears unrealistic near the hot, horizontal plate in
convection where vertical gradients are predominant when vertical
rising structures like plumes are present.

In this study we propose and demonstrate a simple, new criterion which
is able to extract line plumes using just the horizontal velocity
field in a plane parallel and close to the hot plate in turbulent
convection for a Rayleigh number range of
$7.7\times~10^5\le Ra_w\le 1.43\times~10^9$ and a Prandtl number range
of $3.96<Pr<5.3$ where, $Ra_w=g\beta \Delta T_w H^3/\nu\alpha$ and
$Pr=\nu/\alpha$ with $\Delta T_w$ being the temperature drop near the
plate; $\Delta T_w$ is half $\Delta T$, the temperature drop across
the fluid layer height $H$, $\nu$ and $\alpha$ the coefficients of
kinematic viscosity and molecular diffusivity, $g$ the acceleration
due to gravity and $\beta$ the coefficient of thermal expansion. The
proposed criterion of negative values of horizontal divergence of the
horizontal velocity field has the advantages that it is simple to
compute from velocity fields in a plane parallel and close to the
plate, it does not need the vertical gradients of velocities, it needs
only the spatial distribution of horizontal velocity components, it
does not need temperature fields and that it does not need any
arbitrary threshold value that varies from flow to flow.  Such a
criterion will be very useful in detecting line plumes in PIV studies
of the velocity field near the plate in turbulent convection. The
paper is organised as follows, we first describe the setup and PIV
diagnostics used to obtain the velocity field in a horizontal plane
close to the hot plate in \S\ref{sec:experiment}. The criteria used in
wall shear turbulence is then applied to this velocity field to show
their inadequacy in \S\ref{sec:appl-crit-from}. The new criterion is
then proposed based on physical considerations in
\S\ref{sec:phys-just} and then verified in
\S\ref{sec:verif-overl-crit}. 

\section{Experiment}
\label{sec:experiment}
\subsection{Apparatus}
\label{sec:apparatus}
\begin{figure}
\centering
\begin{subfigure}[t]{0.03\textwidth}
(a)
\end{subfigure}
\begin{subfigure}[t]{0.52\textwidth}\caption{}\label{fig:Expt}\vspace{-1cm}
  \includegraphics[width=\linewidth,valign=t]{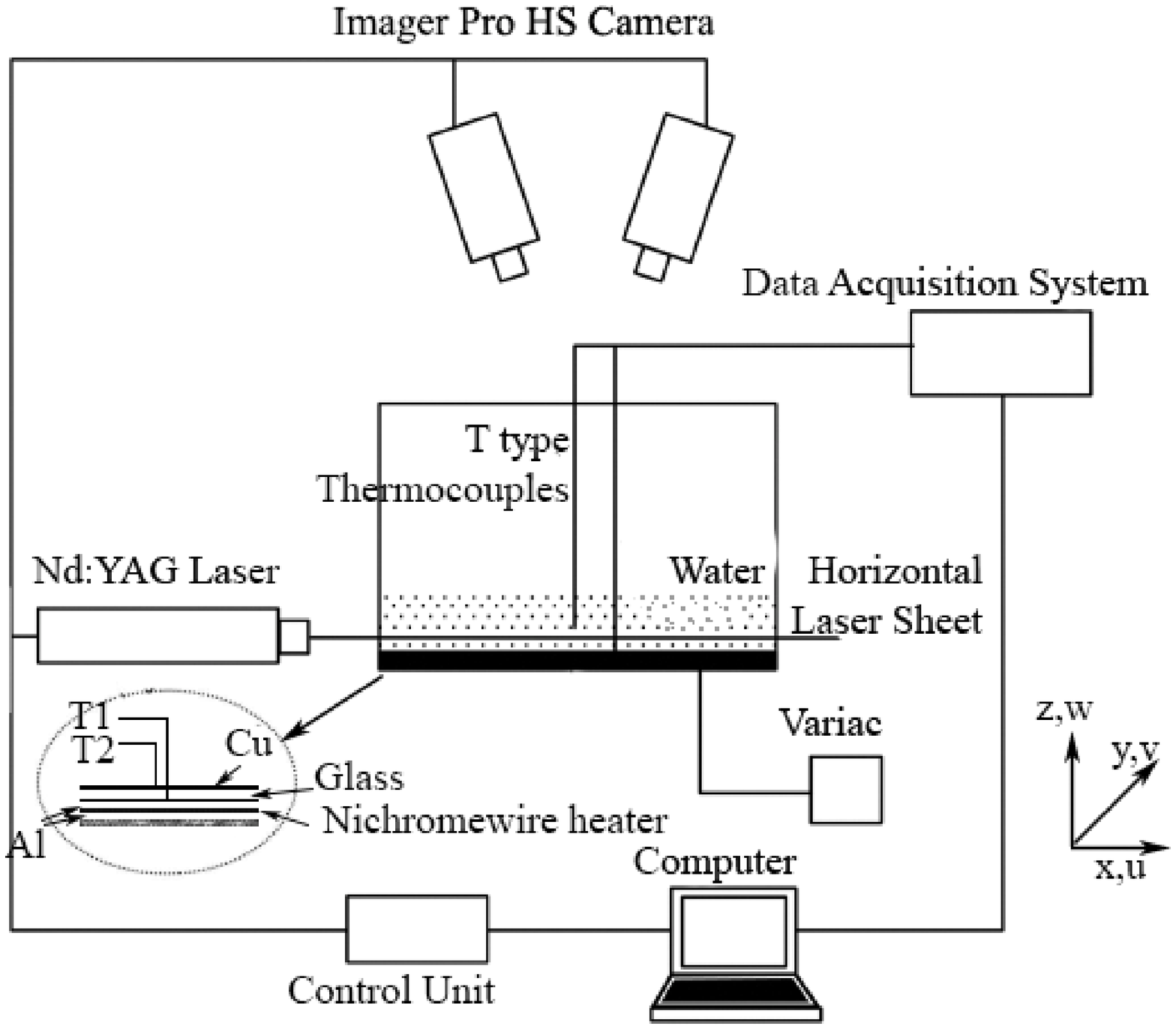}
\end{subfigure}\hfill
\begin{subfigure}[t]{0.03\textwidth}
(b)
\end{subfigure}
\begin{subfigure}[t]{0.4\textwidth}\caption{}\label{fig:dye}\vspace{-0.75cm}
\includegraphics[width=\linewidth,valign=t]{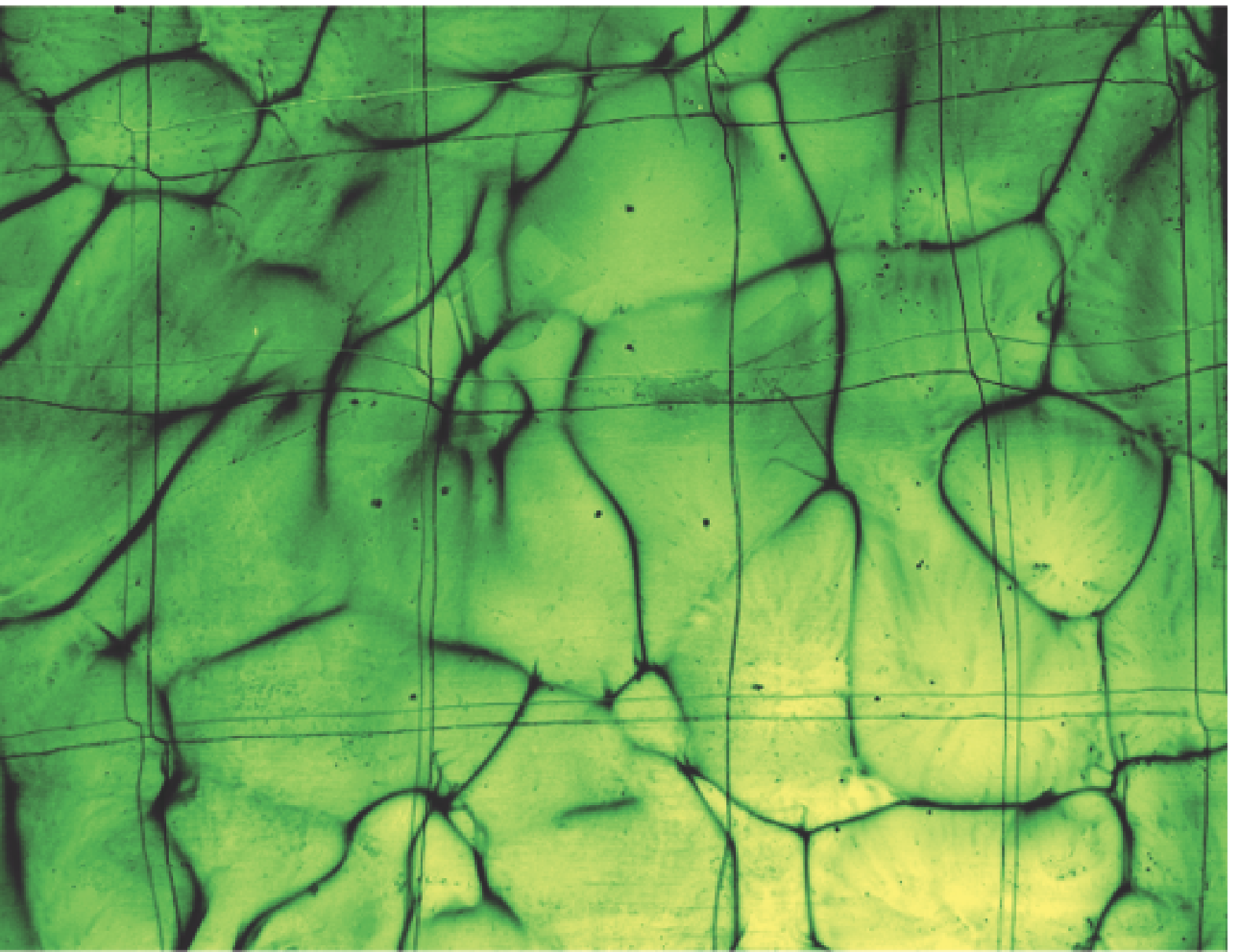}
\end{subfigure}
\caption[]{(a), Schematic of the present experimental setup; (b), Line
  plumes at $Ra_w=7.5\times10^5$ detected by electrochemical dye
  visualization~\cite*[][]{Puthenveettil11}.}
\label{fig:setupvis}
\end{figure}

The velocity field measurements were conducted for unsteady
temperature driven convection in water in an open top glass tank of
cross sectional area $300\times300$ mm$^2$ and height of 250 mm
insulated on the sides as shown in Figure~\ref{fig:Expt}. 
The constant heat flux through the bottom, horizontal, copper plate and
the water layer height $H$ were changed to achieve the range of $Ra_w$
and $Pr$ shown in table~\ref{tab:exppara}. The heat flux was provided
by a nichrome wire heater sandwiched between two aluminum plates and
connected to a
variac. 
The temperature difference $(T_1-T_2)$ across a glass plate, placed
between the upper aluminum plate and the copper plate was measured
using 
T-type thermocouples to estimate the heat
flux. 
The temperature difference between the bottom plate and the bulk
$T_w-T_B=\Delta T_w$
was measured using thermocouples kept touching the copper plate and in
the bulk fluid. 
Measurements were made only after a quasi-steady state was
attained after around 3 hrs, in which $\Delta T_w$ remained a
constant, as monitored by 
a data logger (Agilent, model 34970A). For more details of the setup,
the reader is referred to
\cite{Guna14}. 
\subsection{Diagnostics}
\label{sec:diagnostics}
\begin{table} 
  \begin{center}
    \begin{tabular}{cccccccccccccccc}
      $Ra_w$ & $\Pran$ & $\Delta T_w $ & $T$&$\nu$ &$\alpha$& $H$
      &$\delta_{pb}$ &$\delta_{nc}$&$h_m$&$A_i$& $t_p$ &$\Delta t$&$D_I$&over-&$L_v$\\$\times
      10^7$&&&& $\times 10^{-7}$&$\times
                                  10^{-7}$&&&&&&&&&lap& \\
             && $^\circ$C& $^\circ$C&
                                      $\mathrm{m}^2\mathrm{s}^{-1}$&
                                                                     $\mathrm{m}^2\mathrm{s}^{-1}$&mm
                                                            &mm&mm&mm&mm$^2$&mm&s&pix&\%&mm\\[3pt]
      $0.0766$ & 5.3     &   0.26   & 30   & 7.99  & 1.502  & 50             & 33.68	& 4.8           & 3.5  & $80\times 80$    & 9.05			& 0.2      & 32	& 0            &  1.27    \\                  
      $0.384$ & 5.19    &   1.26   & 31   & 7.82  & 1.508  & 50             & 23.29          & 2.75          & 2 & $145\times 110$  & 7.93			& 0.2      & 32    & 50           &  2.12        \\
      $3.40$ & 4.96    &   1.30   & 33   & 7.51  & 1.51   & 100            & 14.08          & 2.6           & 2  & $85\times 65$    & 7.75			& 0.1      & 32     & 50           &  1.08      \\
      $7.29$ & 4.75    &   2.6    & 35   & 7.22  & 1.518  & 100            & 11.70          & 1.97          & 1.5 & $85\times 65$    & 7.05			& 0.1      & 32     & 50           &  1.08      \\
      $13.2$ & 4.28    &   4      & 40   & 6.57  & 1.53   & 100            & 9.88           & 1.53          & 1.2 & $85\times 65$    & 6.24			& 0.1      & 32     & 50           &  1.08       \\
      $58.1$ & 4.86    &   2.68   & 34   & 7.36  & 1.51   & 200            & 7.41           & 1.99          & 1.4 & $145\times 110$  & 7.06			& 0.0667   & 64     & 75           &  1.8      \\
      $143$ & 3.96    &   4.8    & 44   & 6.11  & 1.54   & 200            & 5.64           & 1.32          & 1.2 & $145\times 110$  & 6.03			& 0.0625   & 64     & 75           &  1.8     \\
  \end{tabular}
  \caption{Values of experimental parameters, length scales and PIV
    parameters. $T$
    is the temperature of the bulk fluid, $D_I$
    is one side of the square interrogation window. }
\label{tab:exppara}
  \end{center}
\end{table}  

The velocity field in a horizontal (x-y) plane, parallel and close to
the hot copper plate at the bottom was obtained by stereo PIV at all
$Ra_w$,
except at $Ra_w=7.66\times10^5$,
where 2D PIV was used. The copper plate was painted black to prevent
reflection of light from its surface. As could be seen from
table~\ref{tab:exppara}, the height of the measurement plane above the
hot plate $(h_m)$
was chosen to be within the Prandtl-Blasius velocity boundary layer
thickness ($\delta_{pb}$)
and within the natural convection velocity boundary layer thickness
($\delta_{nc}$),
estimated from~\cite{Ahlers09} and~\cite{Puthenveettil11}.
The flow was seeded with poly-amide spheres (mean diameter $d_p=55$
$\mu$ m and density $\rho_p=1.012$ g cm$^{-3}$), which were
illuminated by a laser sheet of thickness $1$~mm from a Nd:YAG laser
(Litron, 100 mJ/pulse). The value of the particle Stokes number, the ratio of
particle relaxation time and the characteristic time scale of the
flow,
was less than $0.0024$; 
the particles hence followed the flow. Two Imager Pro HS (LaVision
GmbH) cameras with $1024\times 1280$ pixel resolution (Imager Pro with
resolution of $2048\times 2048$ pixels for the 2D2C experiment) 
were used to capture the particle images of area $A_i$ at the center
of the plate so that $A_i$ had at least $8$ to $12$ plumes; the values
of $A_i$ are shown in table~\ref{tab:exppara}. The cameras with the
Scheimpflug adapters were kept at an inclination of $25^\circ$ with
the $z$ axis, with the depth of field at around $2$ mm, greater than
the laser sheet thickness. A single pulse, single frame mode was used
to capture the particle images with the 
laser pulse separation $\Delta t$ chosen such that the highest out of
plane particle displacement, due to the center line plume velocity,
was not more than one fourth of the laser sheet thickness. The maximum
in-plane particle displacement was around 10 pixels at each
$Ra_w$. 

A high pass filter was applied on the particle images, in order to
make the varying background uniform. A stereo cross correlation method
which calculates a 2D-3C vector field was used to evaluate the vector
field. The size of the interrogation window was chosen so that there
was a minimum of three velocity vectors within the plume thickness
$t_p$ at any $Ra_w$ where, $t_p$ was estimated from the similarity
solutions for line plumes by~\cite{Gebhart70}. 
Table~\ref{tab:exppara} shows the values of $t_p$ and $L_v$, the
spatial resolution of velocity vectors at different $Ra_w$; the number
of vectors in $t_p$ varied from about seven at the lowest $Ra_w$ to
about three at the highest $Ra_w$. The size of the interrogation
window was also limited by the condition that the displacement due to
the largest velocity in the horizontal plane was less than one fourth
the interrogation window size and that a minimum of ten particles were
present in the interrogation window for robustness of correlation.
The bias errors
were reduced by using a multipass adaptive
window cross-correlation technique.
Sub-pixel interpolation ensured that the peak lock value
was less than $0.1$, the acceptable limit of peak
locking effect.  The spurious
vectors were removed by applying a median filter of
$3 \mathrm{pix}\times 3\mathrm{pix}$ neighbourhood with interpolated
vectors replacing the spurious vectors. 
A $3 \mathrm{pix}\times 3\mathrm{pix}$ smoothing filter was applied to
the final vector field to reduce noise. Such smoothed images were used
for the calculation of derivatives. 
\section{Extraction of line plumes from the velocity field}
\label{sec:extr-line-plum}
Figure~\ref{fig:dye} shows an example of the pattern of line plumes
obtained by electrochemical dye visualization~\cite*[][]{baker66}
using Thymol blue at $Ra_w= 7.5\times
10^5$~\cite*[][]{Guna14}.
Application of any criterion to detect coherent structures in
turbulent convection should be first able to qualitatively reproduce
the connected, line type of plume structure similar to that in
figure~\ref{fig:dye}. 
\subsection{Application of criteria from wall turbulence}
\label{sec:appl-crit-from}
\begin{figure}
\centering
\begin{subfigure}[t]{0.02\textwidth}
(a)
\end{subfigure}
\begin{subfigure}[t]{0.46\textwidth}\caption{}\label{fig:Qcrit}\vspace{-1cm}
\includegraphics[width=\linewidth,valign=t]{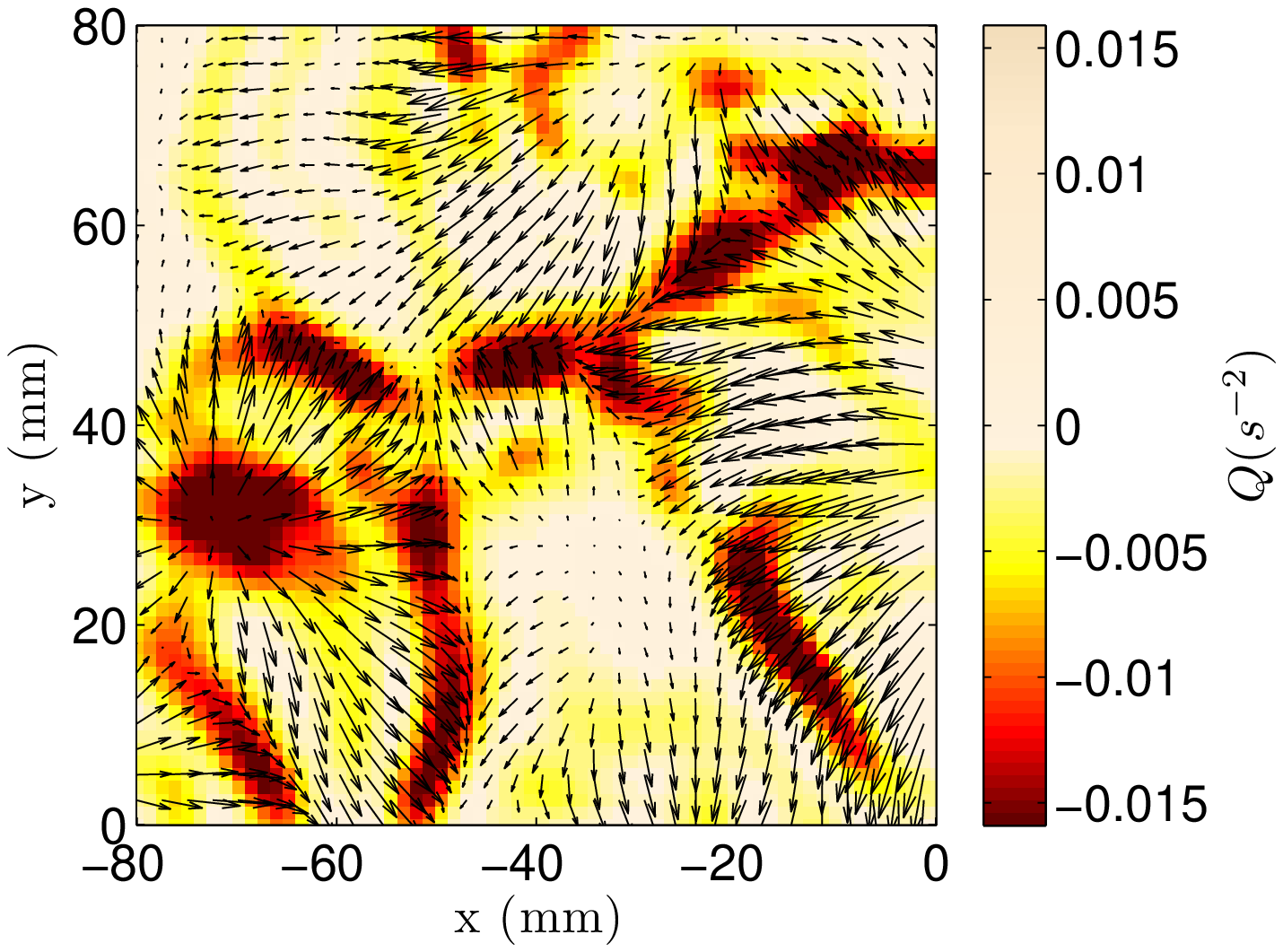}
\end{subfigure}\hfill
\begin{subfigure}[t]{0.02\textwidth}
(b)
\end{subfigure}
\begin{subfigure}[t]{0.46\textwidth}\caption{}\label{fig:lambda2}\vspace{-1cm}
\includegraphics[width=\linewidth,valign=t]{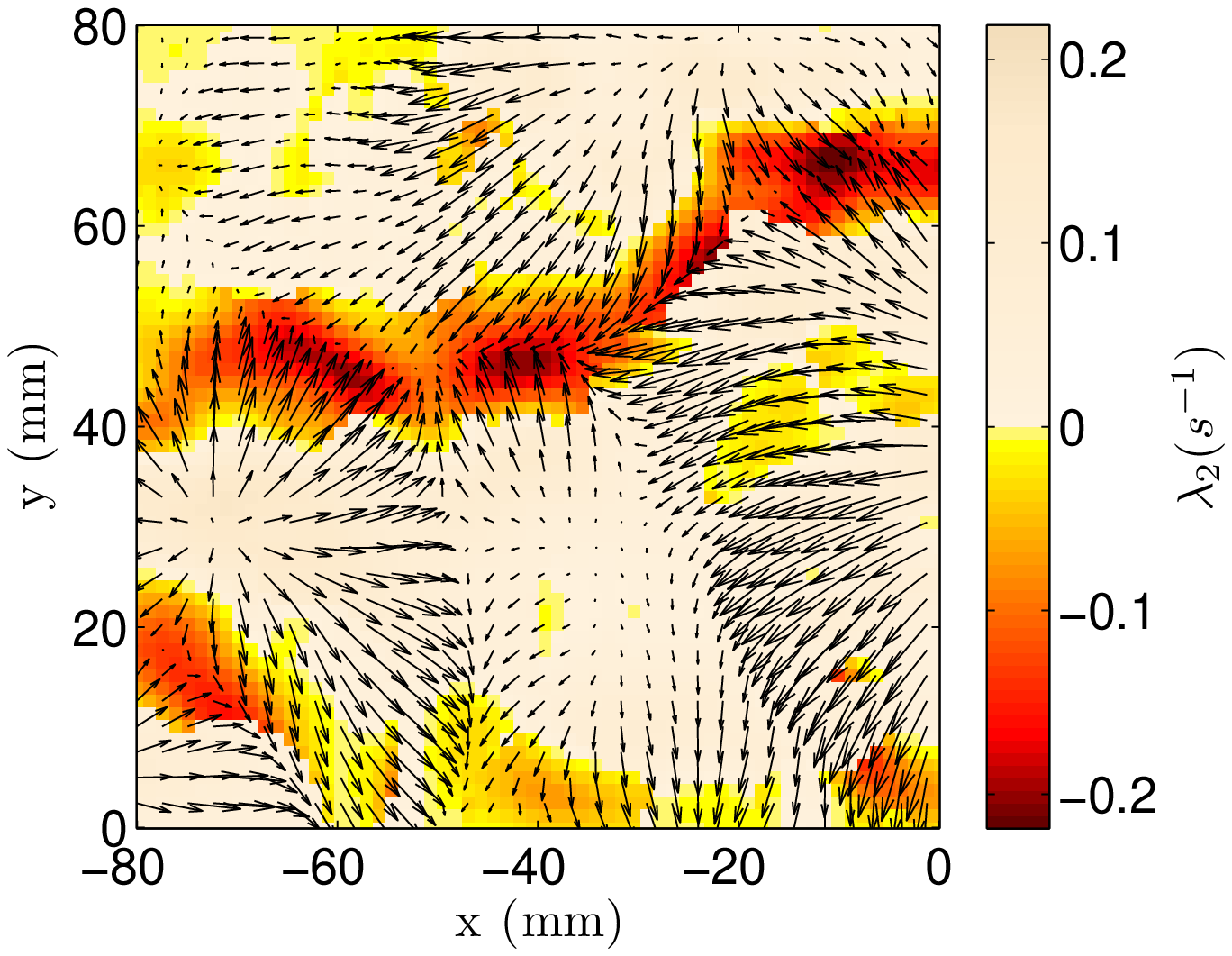}
\end{subfigure}\\
\begin{subfigure}[t]{0.02\textwidth}
(c)
\end{subfigure}
\begin{subfigure}[t]{0.46\textwidth}\caption{}\label{fig:lambdaci}\vspace{-1cm}
\includegraphics[width=\linewidth,valign=t]{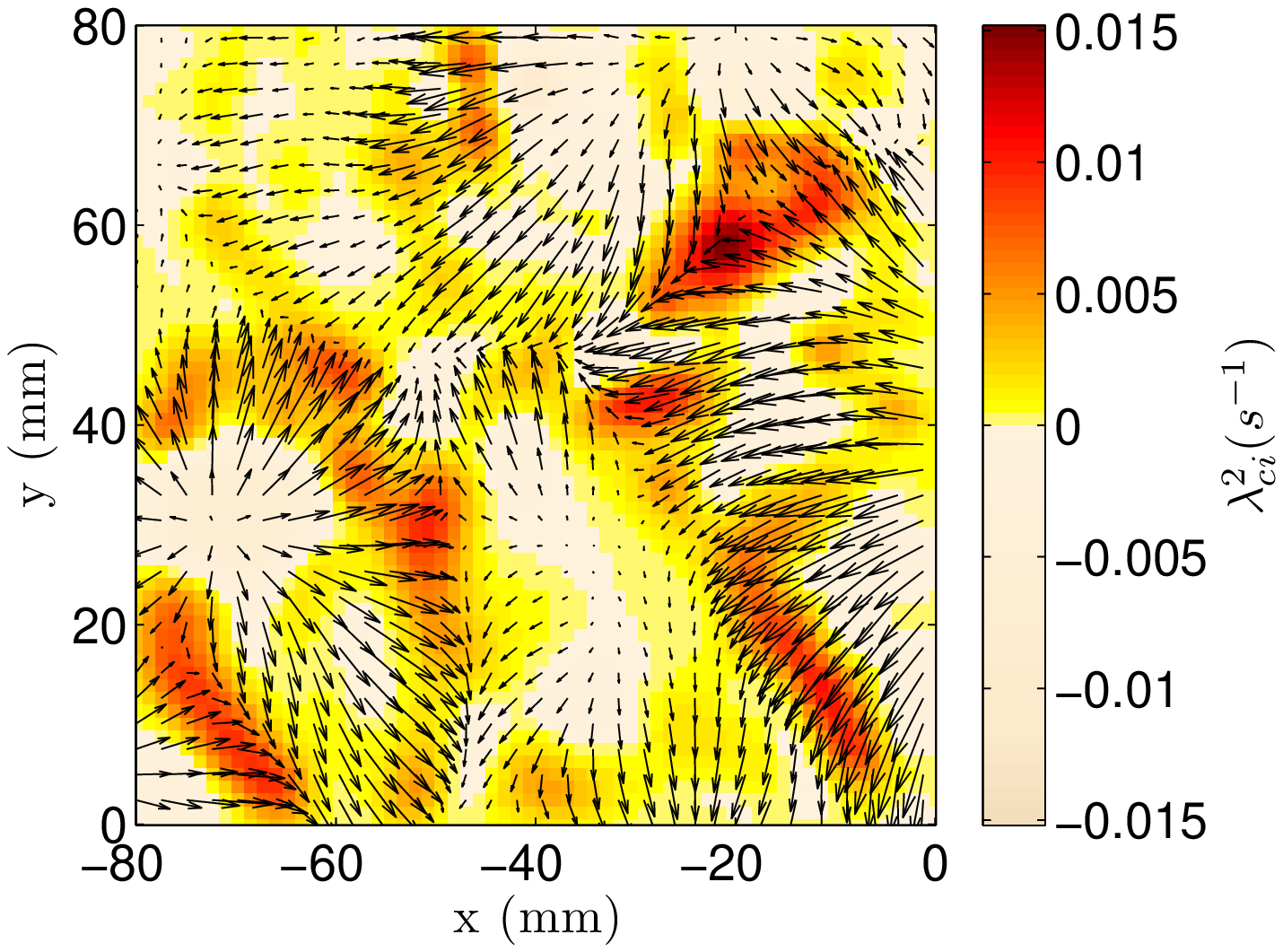}
\end{subfigure}\hfill
\begin{subfigure}[t]{0.02\textwidth}
(d)
\end{subfigure}
\begin{subfigure}[t]{0.46\textwidth}\caption{}\label{fig:delta2u}\vspace{-1cm}
\includegraphics[width=\linewidth,valign=t]{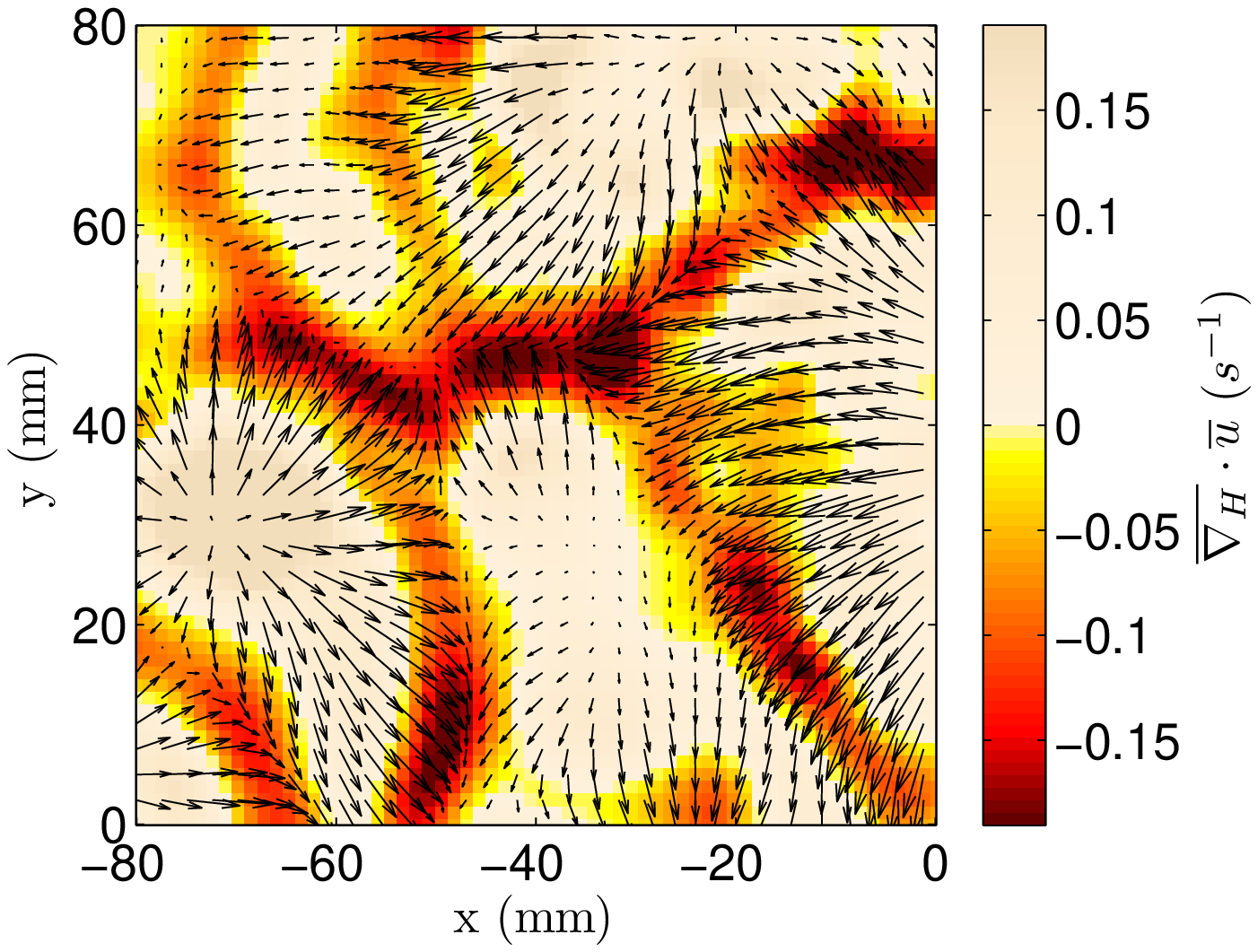}
\end{subfigure}
\caption[]{Coherent structure extraction methods applied on a vector
  field at $Ra_w=7.66\times10^5$; (a), Q-criterion; (b), $\lambda_2$
  criterion; (c), $\lambda_{ci}$ method and (d),
  $\overline{\nabla_H}\cdot\overline{u}$ criterion.}
\label{fig:ext_criteria}
\end{figure}
We first apply the $Q$-criterion~\cite*[][]{hunt88:_eddies},
$\lambda_2$ criterion~\cite*[][]{Jeong95} and the swirling strength
($\lambda_{ci}$) criterion~\cite*[][]{Zhou99} on our velocity field to
see whether the line plume structure is captured by these criteria. In
all cases, the terms involving vertical velocity gradients are set to
zero. The second invariant of $\overline{\nabla} \overline{u}$,
$Q=\frac{1}{2}(\|\Omega\|^2-\|S\|^2)$, where $\|\Omega\|$ and $\|S\|$
are the symmetric and the anti-symmetric components of
$\overline{\nabla} \overline{u}$ respectively, is a local measure of
excess rotation rate relative to the strain rate. Regions with $Q>0$
denote vortices and regions with $Q<0$ denotes regions of larger
strain rate. Figure~\ref{fig:Qcrit} shows the spatial map of $Q$,
calculated from our 3D PIV vector field in a horizontal plane close to
the plate, overlaid over the vector field at $Ra_w =7.66\times 10^5$.
The regions of $Q<0$ captures line-like regions, into which the flow
converges locally, which could be the plumes. However, $Q<0$ regions
also include regions from which the flow diverges, namely regions at
the center in between the line-like regions, see at (-70, 30) and
(-20,70) in figure~\ref{fig:Qcrit}, thereby making this criterion
unsuitable for extracting plumes.

In $\lambda_2$ criterion, the local pressure minimum within a vortex
core is identified as the regions with two negative eigen values of
the symmetric tensor $\xi=\|\Omega\|^2+\|S\|^2$. Since the eigen
values of $\xi$ are real, from the ordered triad of the three local
eigen values $\lambda_1\ge \lambda_2\ge \lambda_3$, the regions with
negative $\lambda_2$ are identified as the vortex cores. The
$\lambda_2$ criterion looks for the excess rotation rate relative to
the strain rate in one plane alone while the $Q$ criterion looks for
it in all directions.
Figure~\ref{fig:lambda2} shows the distribution of $\lambda_2$
calculated from the vector field in figure~\ref{fig:Qcrit}, overlaid
over the same vector field.
Far less line type structures are captured by the $\lambda_2$
criterion, compared to that by the $Q$ criterion, possibly because the
terms set to zero in $\xi$ are important in the present flow.

Swirling strength ($\lambda_{ci}$) is the imaginary part of the
complex eigen value of the velocity gradient tensor
$\overline{\nabla} \overline{u}$, which is a measure of the local
swirling rate inside a vortex. 
As shown in figure~\ref{fig:lambdaci}, distribution of the positive
values of $\lambda_{ci}^2$ captures the line type structures. However,
it also captures broad patches in various regions, like in locations
(-60, 20), (-40, 5), (-25, 75) and (0, 40) in
figure~\ref{fig:lambdaci}, which do not seem to be plumes. It hence
appears that $Q$, $\lambda_2$ and $\lambda_{ci}$ criteria are not able
to isolate plumes alone, mostly due to the non-availability of the
vertical gradients of velocity, which seems to be important to
correctly estimate the values of $Q$, $\lambda_2$ and $\lambda_{ci}$
for the present flow. Since most PIV studies have the velocity field
only in a horizontal plane and not in a 3D volume, as is often the
case in numerical simulations where these criteria are more
successful, these criteria can hence not be used to detect line plumes
from PIV studies of turbulent convection. We now propose a new and
simple criterion to detect line plumes from velocity fields in a plane
parallel and close to the convecting surface.
\subsection{$\overline{\nabla_H}\cdot\overline{u}$ criterion}
\label{sec:overl-crit}
\subsubsection{Physical justification}
\label{sec:phys-just}
\begin{figure}
  \centering
  \includegraphics[width=0.6\textwidth]{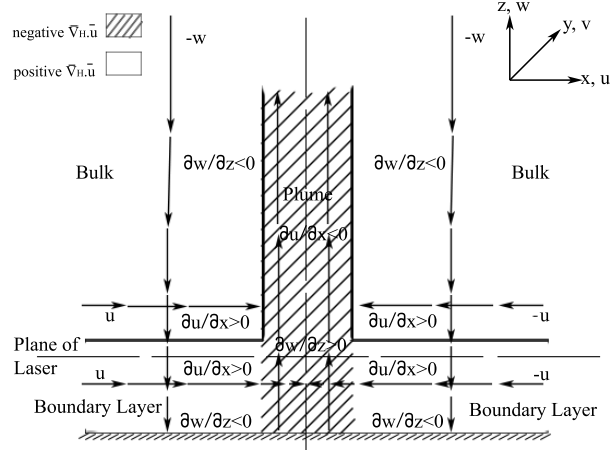}
\caption{Schematic of the velocity variation in the plume, the bulk
  and the boundary layer.}
\label{fig:velvarsch}
\end{figure}

Plumes, shown as the hatched region in figure~\ref{fig:velvarsch}, are
known to have positive vertical spatial acceleration,
$\partial w/\partial z>0$~\cite*[][]{Gebhart70}. At the same time, due
to symmetry, horizontal velocities within the plumes decrease as we
approach the center of the plumes from its edges, resulting in
negative $\partial u/\partial x$ in the hatched region of
figure~\ref{fig:velvarsch}. Plumes do not occur in isolation, but
cause flows around them due to the ambient fluid being entrained into
them. Such entrainment flows are expected to have negative
$\partial w/\partial z$ 
since the large scale flow in the bulk is of an order higher in
magnitude than the entrainment flows into the plumes and the boundary
layers~\cite*[][]{Guna14}. Hence, $\partial w/\partial z$ is expected
to be negative in the regions shown as bulk in
figure~\ref{fig:velvarsch}. In addition, the presence of plumes also
results in increasing horizontal velocities in the bulk
($\partial u/\partial x>0$) due to the entrainment, as shown by the
horizontal arrows in the bulk region in figure~\ref{fig:velvarsch}.
If we now consider regions within the boundary layers feeding the
plumes on either sides at the bottom of the plumes, if these are
natural convection boundary layers, the horizontal velocities are
expected to increase towards the plumes since the driving horizontal
pressure gradient increases with increase in boundary layer thickness
($\Delta p \approx \rho g \beta\Delta T \delta_{nc}$) resulting in
$\partial u/\partial x >0$, as shown in the boundary layer regions in
figure~\ref{fig:velvarsch}. Hence, we see that the flow near the plate
separates into two regions, viz. (a)~the plumes where
$\partial w/\partial z>0$ and $\partial u/\partial x<0$ and (b)~the
bulk and the boundary layers where $\partial w/\partial z<0$ and
$\partial u/\partial x>0$; these two regions are shown as the hatched
and the unhatched regions respectively in
figure~\ref{fig:velvarsch}. We hence need a criterion that
distinguishes the $\partial w/\partial z>0$, $\partial u/\partial x<0$
regions as the plumes from the $\partial w/\partial z<0$,
$\partial u/\partial x>0$ regions as the bulk or the boundary layer.

Horizontal, two dimensional divergence of the horizontal velocity
components, calculated in a plane parallel and close to the plate,
\begin{equation}
  \label{eq:delhu}
  \overline{\nabla_H}\cdot\overline{u}= \frac{\partial u}{\partial x} +\frac{\partial v}{\partial y}, 
\end{equation}
seems to satisfy the above mentioned requirements, with the subscript
$H$ indicating that the divergence operator is applied only on the
horizontal ($x$ and $y$) components of the velocity field, namely $u$
and $v$.  Since the flow is incompressible,
$\overline{\nabla_H}\cdot\overline{u} = -\partial w/\partial z $;
negative values of $\overline{\nabla_H}\cdot\overline{u}$ implies that
such regions will have positive $\partial w/\partial z$ and vice
versa. Further, $\overline{\nabla_H}\cdot\overline{u} <0$ implies that
these regions are also the regions in which the horizontal velocities
reduce in magnitude along the direction of these velocities
(figure~\ref{fig:velvarsch}). As we saw earlier, such regions are
likely to be the plumes. Similarly, regions with
$\overline{\nabla_H}\cdot\overline{u} >0$ will have increasing
horizontal velocities in the direction of these velocities as well as
$\partial w/\partial z <0$; such regions are more likely to be the
ambient fluid or the boundary layer regions in between the plumes.
\begin{figure}
\centering
\begin{subfigure}[t]{0.02\textwidth}
(a)
\end{subfigure}
\begin{subfigure}[t]{0.47\textwidth}\caption{}\label{fig:divra6}\vspace{-1cm}
  \includegraphics[width=\linewidth,valign=t]{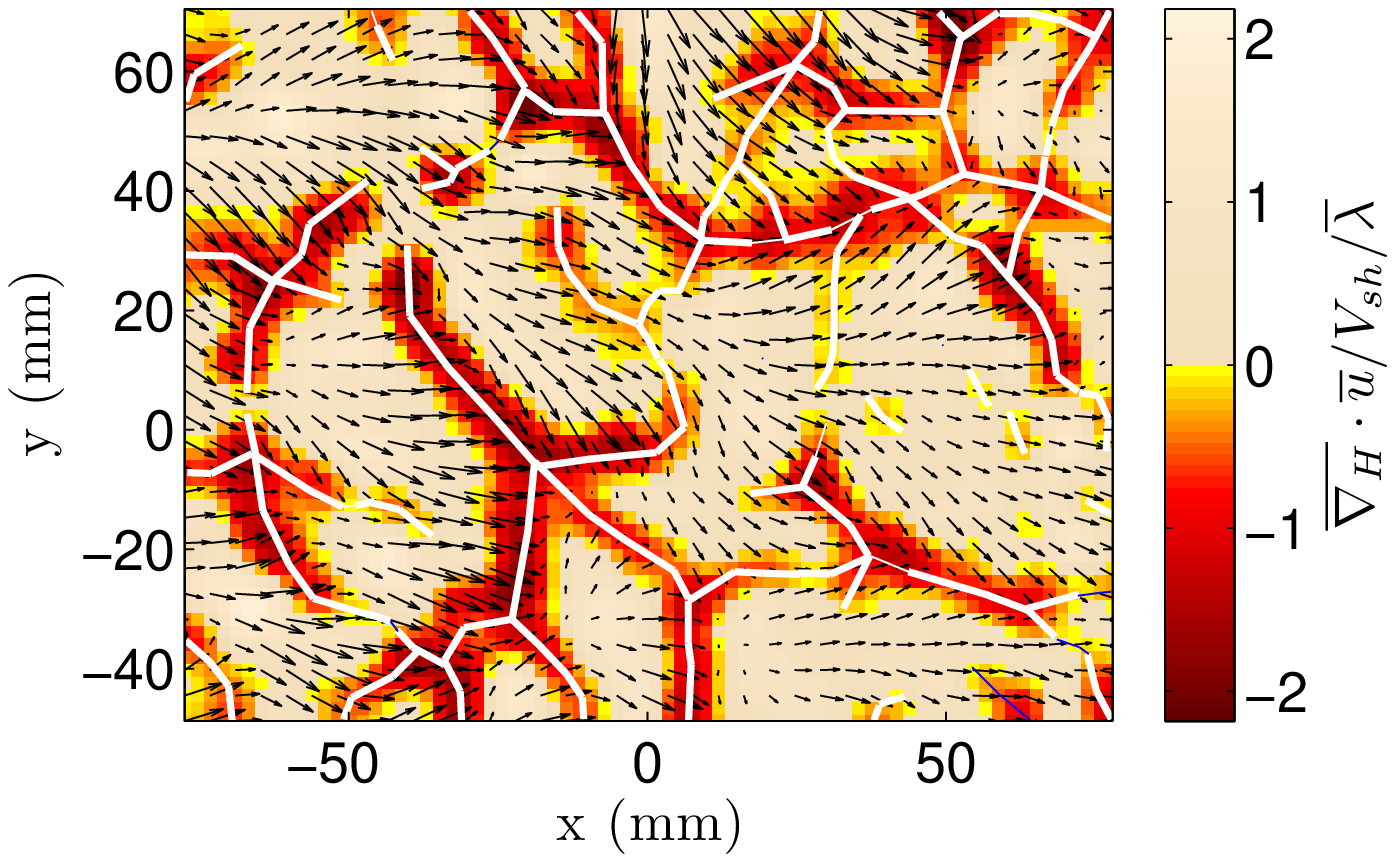}
\end{subfigure}\hfill
\begin{subfigure}[t]{0.02\textwidth}
(b)
\end{subfigure}
\begin{subfigure}[t]{0.47\textwidth}\caption{}\label{fig:divra9}\vspace{-1cm}
  \includegraphics[width=\linewidth,valign=t]{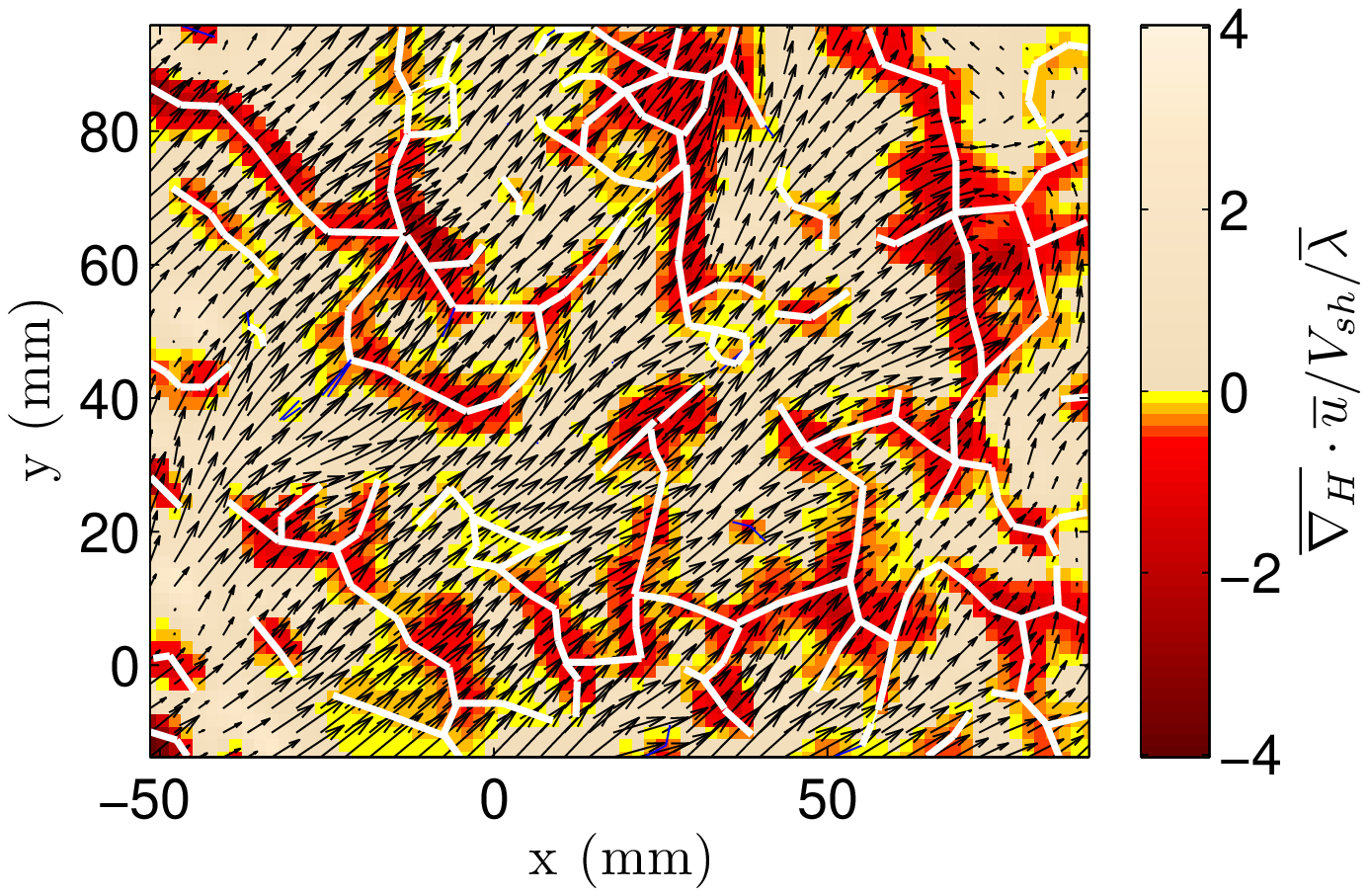}
\end{subfigure}
\caption{$\overline{\nabla_H}\cdot\overline{u}$ overlaid over the 2D
  velocity field covered with short linear segments drawn over
  negative divergence regions at (\textit{a}) $Ra_w=3.84\times10^6$,
  Pr=5.2 and (\textit{b}) $Ra_w=1.43\times10^9$, Pr=3.96.}
\label{fig:div}
\end{figure} 
\subsubsection{Verification of the
  $\overline{\nabla_H}\cdot\overline{u}$ criterion}
\label{sec:verif-overl-crit}
We now apply this simple criterion to the horizontal velocity field
obtained close to the plate in our experiments, on which the other
criteria was applied in
\S~\ref{sec:appl-crit-from}. Figure~\ref{fig:delta2u} shows the
instantaneous horizontal velocity field overlaid over the distribution
of $\overline{\nabla_H}\cdot\overline{u}$ calculated from the same
velocity field using central differencing at $Ra_w=7.66\times 10^5$.
Only the variations of negative $\overline{\nabla_H}\cdot\overline{u}$
values are shown in dark color while all the positive values are set
to a uniform grey color. Line like regions, similar to plumes, are
detected by the $\overline{\nabla_H}\cdot\overline{u}$
criterion. Unlike in the case of application of the $Q$ criterion in
figure~\ref{fig:Qcrit}, the central, diverging flow regions between
the plumes are not picked up by the
$\overline{\nabla_H}\cdot\overline{u}$ criterion.  In comparison to
the application of $\lambda_2$ criterion in figure~\ref{fig:lambda2},
far more line like regions are picked up by the
$\overline{\nabla_H}\cdot\overline{u}$ criterion. A comparison of
figures~\ref{fig:lambdaci} and~\ref{fig:delta2u} shows that
$\lambda_{ci}$ picks up more line like regions than the
$\overline{\nabla_H}\cdot\overline{u}$ criterion, from which to
identify the plumes the $\lambda_{ci}$ criterion needs an unknown
value of threshold for $\lambda_{ci}$. Hence, the negative values of
$\overline{\nabla_H}\cdot\overline{u}$ seems to fall on line like
regions alone, which seems to be plumes, with no need for an arbitrary
and a priorly unknown threshold value for the criterion. 

We now see whether the behaviour of such line like regions are
qualitatively similar to that of line plumes. We define the
dimensionless, two dimensional, horizontal divergence of the velocity
field as,
\begin{equation}
  \label{eq:dimless2ddiv}
  \zeta=\frac{\overline{\nabla_H}\cdot\overline{u}}{V_{sh}/\overline{\lambda}}
\end{equation}
where $V_{sh} $ is the large scale velocity obtained from the shear
Reynolds number, $Re_{sh} = V_{sh}H/\nu = 0.55Ra_w^{4/9}Pr^{-2/3}$,
given as (5.2) in~\cite{Guna14} and
\begin{equation}
  \label{eq:lambda}
\overline{\lambda}=C_1Z_w Pr^{n_1}  
\end{equation}
is the mean plume spacing with $C_1=47.5$, $n_1=0.1$ and
$Z_w=\left(\nu\alpha/g\beta \Delta T_w\right)^{1/3}$ is the length
scale near the plate due to the balance of driving buoyancy forces and
the dissipative effects; see~\cite{Puthenveettil11} for the physical
significance of $Z_w$.

Figure~\ref{fig:div} shows the distribution of the negative values of
$\zeta$ in dark color overlaid over the velocity vector field at
$Ra_w=3.84\times10^6$ and $Ra_w=1.43\times10^9$. Since the maximum
value of $\zeta$ is of order one, $V_{sh}/\overline{\lambda}$ is the
appropriate characteristic divergence of the velocity
field. Figure~\ref{fig:divra9} shows that an increase in $Ra_w$
increases the length of the negative $\zeta$ regions, as expected for
line plumes~(\cite{Puthenveettil11}). At the lower $Ra_w$ in
figure~\ref{fig:divra6}, the negative $\zeta$ regions have flows going
into them, as would be expected for plumes, since feeding of plumes by
the boundary layers on their sides is predominant in the absence of
strong large scale flows at lower
$Ra_w$~\cite*[][]{Guna14}. Figure~\ref{fig:div} shows that, similar to
line plumes, these regions also align along the predominant large
scale flow direction in regions of stronger
shear~\cite*[][]{Puthenveettil05b}. These qualitative considerations
strongly suggest that the negative $\zeta$ regions are the line
plumes.
\begin{figure}
  \centering
  \includegraphics[width=0.6\textwidth]{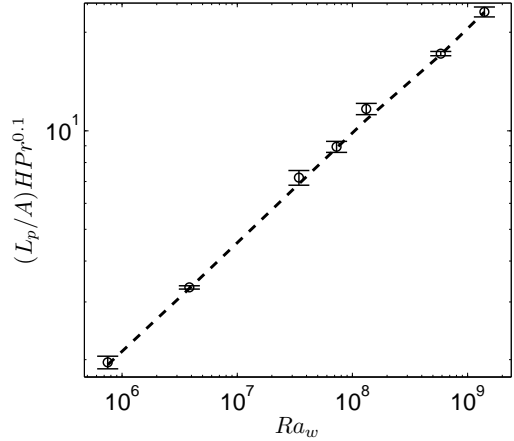}
  \caption{Comparison of the length of the negative
    $\overline{\nabla_H}\cdot\overline{u}$ regions ($\circ$) with the
    plume length given by (\ref{eq:Length}) ($---$) for the range of
    $Ra_w$ and $Pr$ in the present experiments.}
\label{fig:plu_length}
\end{figure}

To quantitatively verify whether the $\zeta<0$ regions are plumes, we
now compare the total length of the negative $\zeta$ regions with the
prediction of the total plume length in an area $A$ of the hot plate,
\begin{equation}
  \label{eq:Length}
  L_p=A/\overline{\lambda}, 
\end{equation}
by~\cite{Puthenveettil11}, where $\overline{\lambda}$ is given by
(\ref{eq:lambda}). Equation~\ref{eq:Length} has been verified with
measurements of plume lengths over six decades of $Ra_w$ and three
decades of $Pr$ by~\cite{Puthenveettil11}. The length of the negative
$\zeta$ regions were measured as in~\cite{Puthenveettil11}, by
covering these regions with short linear segments as shown by the
white lines in figure~\ref{fig:div}; the sum of the length of these
segments gives the total length of the negative $\zeta$ regions.
Figure~\ref{fig:plu_length} shows the total length of the negative
$\zeta$ regions plotted along with (\ref{eq:Length}) for the range of
$Ra_w$ in the present study. The error bars show the range of data
from measurements at different instants at the same $Ra_w$. The
excellent match between the theoretical prediction of $L_p$ and the
length of negative $\zeta$ regions from the present measurements gives
us confidence that these regions are the plume regions.

We now examine the thickness of the negative $\zeta$ regions vis-a-vis
the plume thickness. A qualitative examination of
figure~\ref{fig:ext_criteria} shows that the thicknesses of the
negative $\overline{\nabla_H}\cdot\overline{u}$ regions are
approximately the same as the plume thicknesses that would be
identified by the other criteria. A quantitative comparison of the
thickness of the plume region, identified by the $\zeta<0$ criterion,
with the plume thickness in an ideal laminar line plume could be
attempted as follows. The horizontal and vertical velocities in a
laminar line plume are,
\begin{equation}
  \label{eq:plumeeqn}
  u=4^{3/4}\frac{\nu}{z}Gr_z^{1/4}\left(\frac{3}{5}f'(\eta)-\frac{2}{5}\eta
    f(\eta)\right),\:w=4^{1/5}{\left(\frac{g\beta Q}{C_p I}\right)}^{2/5}{\left(\frac{z}{\mu
        \rho}\right)}^{1/5} f'(\eta),
\end{equation}
where, $Q$ is the heat flux per unit length of the line source, $C_p$
is the specific heat at constant pressure,
$I =\int_\infty^\infty f'\phi d\eta$ with
$\phi=(T-T_\infty)/(T_0-T_\infty)$ where $T$ is the temperature with
subscripts $0$ and $\infty$ denoting plume center line and ambient,
and the similarity variable $\eta={(x/z)(Gr_z/4)}^{1/4}$ with
$Gr_z=g\beta (T_0-T_\infty)z^3/\nu^2$ being the Grashoff number based
on $z$~\cite*[][]{Gebhart70}. $f$ is the dimensionless stream function
given by $\psi=4\nu{(Gr_z/4)}^{1/4} f(\eta)$, with $\psi$ being the
stream function and $'$ denotes differentiation with respect to
$\eta$.  The horizontal velocity in (\ref{eq:plumeeqn}) shows that
$\partial u/\partial x$ is negative when $\eta>f'(\eta)/2f''(\eta)$
which is always satisfied within a plume since $\eta$ is positive and
$f'(\eta)$ is a positive but a decreasing function with $\eta$. The
vertical velocity in (\ref{eq:plumeeqn}) reveals that, at any height
$z$, $\partial w/\partial z =w/5z=0$ occurs at the same horizontal
location as $w=0$, corresponding to $f'(\eta)=0$; the ideal
theoretical plume thickness calculated from $w=0$ and $\zeta=0$ would
hence give the same value.

A comparison of the plume thicknesses obtained from the $\zeta<0$
criterion with that obtained from a threshold of dimensionless RMS
temperature fluctuations $\sigma=\sqrt{\overline{T'^2}}/\Delta T$ has
recently been attempted by~\cite{gastine15:_turbul_rayleig_benar}. The
plume regions identified by $\overline{\nabla_H}\cdot\overline{u}<0$
coincided with that identified by the $\sigma$ criterion, however the
plume thicknesses based on $\sigma$ increased with decrease in
$\sigma$, with $\sigma=0.25$ being close to, but slightly lower than,
the thicknesses given by $\zeta<0$ criterion. They also showed that
the PDF of the inter plume areas obtained by the $\zeta<0$ criterion
has the same shape as that given by the threshold $\sigma=0.25$, but
is slightly shifted to the left. It hence appears that the thickness
of the plume regions identified by the $\zeta<0$ criterion would be
approximately equal to that estimated by a threshold of
$\sigma\approx 0.2$.
\section{Conclusions}
\label{sec:concl-disc}
In this paper we showed that a simple criterion of
$\overline{\nabla_H}\cdot\overline{u}<0$ picks up the plume regions in
turbulent convection far better compared to the other criteria for
detection of coherent structures from velocity fields alone, proposed
for shear driven wall turbulence. The total length of plumes detected
by such a criterion matched quite accurately with the available
theoretical predictions of~\cite{Puthenveettil11}. The thicknesses of
the plume regions predicted by the criterion was approximately equal
to that obtained by a threshold of
$\sqrt{\overline{T'^2}} =0.2\Delta T$ by
\cite{gastine15:_turbul_rayleig_benar}; the plume regions coincided
from both the criteria. The $\overline{\nabla_H}\cdot\overline{u}<0$
criterion was obtained from a physical picture of the flow induced by
a line plume, where, due to the entrainment of the plume, regions
outside the plume were expected to have increasing horizontal flow
velocities as they approach the plume in a horizontal plane, resulting
in positive values of $\overline{\nabla_H}\cdot\overline{u}$. However,
in the presence of a strong external shear due to the large scale
flow, which the line plumes in turbulent convection are subject to at
high $Ra_w$, such positive spatial accelerations of the horizontal
flows towards the plumes could be affected by the external shear,
thereby affecting the thickness of the plume regions estimated by the
$\overline{\nabla_H}\cdot\overline{u}<0$ criterion. In such
situations, a modification of the present criterion to include the
effect of the ambient might be needed to get an accurate estimate of
the plume thickness. However, such an effect would be negligible when
the plane of analysis becomes close to the hot surface, as is in the
present study, where the effects of large scale flow would be small
even at high $Ra_w$.

\bibliographystyle{jfm}
\bibliography{Koothur_Vipin_1}
\end{document}